\documentstyle{aipproc}

\def\ha{{1\over 2}}
\def\lan{\langle}
\def\ran{\rangle}
\def\bra#1{\lan#1|}
\def\ket#1{|#1\ran}

\def\normord#1{\mathopen{\hbox{\bf:}}#1\mathclose{\hbox{\bf:}}}
\def\frac#1,#2{{#1\over #2}}

\def\fr#1,#2{{#1\over #2}}
\def\e{\epsilon}
\def\ra{\rightarrow}

\begin{document}
\title{1+1 Gauge Theories in the\\ Light-Cone Representation}

\author{Gary McCartor$^*$ and Yuji Nakawaki$^{\dagger}$}
\address{$^*$Department of Physics, SMU, Dallas, Texas 75275\\
$^{\dagger}$Division of Physics and Mathematics, Setsunan University, Osaka 572-8508}

\maketitle

\begin{abstract}
We present a representation independent solution to the continuum Schwinger model in light-cone ($A^+ = 0$) gauge.  We then discuss the problem of finding that solution using various quantization schemes.  In particular we shall consider equal-time quantization and quantization on either characteristic surface, $x^+ = 0$ or $x^- = 0$. 
\end{abstract}

\section*{Introduction}
We shall give a solution to the light-cone gauge Schwinger model in the continuum\cite{nm99}.  In early light-cone meetings many people gave talks on the Schwinger model and a number of papers were published.  All of these solutions made use of periodicy conditions to regulate the infrared singularities of the Schwinger model.  Since a solution with nice periodicity conditions on a space-like surface does not posess nice periodicity conditions on a light-like surface (or even another space-like surface) the comparisons of quantization on a light-like surface with quantization on a space-like surface were never direct.  Since the continuum is the continuum on any surface, in the present paper we shall be able to compare quantizaton proceedures which must arrive at a common solution.

Some of the points we will make about the various cases are as follows:

\noindent For equal-time quantization we find

\begin{flushleft}
\hskip.5truein EASY FORMULATION \\
\hskip.5truein DIFFICULT SOLUTION \\
\hskip.5truein COMPLEX, DYNAMICALLY DETERMINED VACUUM \\
\end{flushleft}

\noindent For light-cone ($x^+ = 0$) quantization we find

\begin{flushleft}
\hskip.5truein MORE DIFFICULT FORMULATION\\
\hskip.5truein EASIER SOLUTION\\
\hskip.5truein VACUUM FIXED BY KINEMATICS AND GAUGE INVARIANCE\\
\end{flushleft}

\noindent For light-cone ($x^- = 0$) quantization (note that in the continuum this is precisely the same as quantizing on $x^+ = 0$ but in the gauge $A^- = 0$ --- the anti-light-cone gauge) we find

\begin{flushleft}
\hskip.5truein UNFAMILIAR FORMULATION\\
\hskip.5truein UNFAMILIAR DEGREES OF FREEDOM\\
\hskip.5truein EASIER SOLUTION\\
\hskip.5truein VACUUM FIXED BY KINEMATICS AND GAUGE INVARIANCE\\
\end{flushleft}

An important point is that even though the vacuum is fixed by kinematics in both the light-cone representations, physical quantities such as the chiral condensate are dynamical just as they are in the equal-time representation.

The solution contains fields which are functions of $x^+$; they therefore appear as dynamical fields when quantized on $t = 0$, zero-mode fields when quantized on $x^+ = 0$ and static fields when quantized on $x^- = 0$.  We shall emphasize the essential nature of these fields to the solution and shall argue that it is natural and not difficult to include them in any of the quantization schemes.

\section*{Light-Cone Gauge Schwinger Model}
The Lagrangian is

$$
 {\cal L} =  i \bar{\psi} \gamma^{\mu}
\partial_{\mu} \psi 
- \fr{1},{4} F^{\mu \nu} F_{\mu \nu} 
-  A^{\mu} {J}_{\mu} -  \lambda A^+
$$
Where $\lambda$ is a Lagrange multiplier field.

The solution is given by

$$
     \Psi_+ = Z_+ e^{\Lambda_+^{(-)}}\sigma_+ e^{\Lambda_+^{(+)}}
$$
$$
      \Lambda_+ = -i2\sqrt{\pi}({\eta}(x^+) + \tilde{\Sigma}(x^+,x^-))
$$
$$
       Z_+^2 = \fr{m^2e^\gamma},{8\pi\kappa}
$$
$$
     \Psi_- = \psi_- = Z_-e^{\Lambda_-^{(-)}}\sigma_- e^{\Lambda_-^{(+)}}
$$
$$
        Z_-^2 = \fr{\kappa e^\gamma},{2\pi}
$$
$$
      \Lambda_- = -i2\sqrt{\pi}\phi(x^+)
$$
$$
      \lambda =m{\partial }_{+}({\eta }-{\phi }) 
$$
$$
        A_+ = \fr{2},{m} \partial_+ ({\eta} + \tilde{\Sigma})
$$

In these equations,$ \tilde{\Sigma}$ is a massive $(\fr{e},{\sqrt{\pi}})$ pseudoscalar; $\phi$ is the $x^+$-dependent piece of a massless scalar  and $\eta$ is the $x^+$-dependent piece of a massless ghost.  These last two fields are regulated with the Klaiber procedure; for example
$$
\phi^{(+)}(x^{+})=i(4\pi )^{-\ha}\int_{0}^{\infty
}dk_{+}k_{+}^{-1}d(k_{+})\left({\rm e}^{-ik_+ x^+}-\theta 
(\kappa -k_{+})\right)
$$

We again wish to emphasize that the above soloution is representation independent.  It is not light-cone quantized or equal-time quantized but is simply the answer.  Any quantization scheme is an attempt to find that answer.

\section*{Role of the Zero-Mode Fields}

We shall refer to the $x^+$-dependent fields as zero-mode fields, although they are such only if quantizing on $x^+ = 0$.  The only operators in the above solution which carry a charge are the spurions, $\sigma_+$ and $\sigma_-$

$$
\sigma_+ ={\rm exp}\Big[ i\sqrt{\pi}\{Q_5 + Q\}(4m)^{-1} +\int_{0}^\kappa dk_1
k_1^{-1}\{\eta(k_1)-\eta^*(k_1)\}\Big]
$$
$$
 \sigma_- ={\rm exp}\Big[ i\sqrt{\pi}\{Q_5 - Q\}(4m)^{-1} + \int_0^{\kappa} dk_1
k_0^{-1}\{d(k_1) - d^*(k_1)\}\Big]  
$$

The charges themselves are made up entirely from operators from the zero-mode fields.  We have

$$
    Q= \int_{-{\infty}}^{\infty}m{\partial}_+({\phi}-{\eta})dx^+        
$$
$$
    Q_5= \int_{-{\infty}}^{\infty}m{\partial}_+({\phi}+{\eta})dx^+
$$

$$
  \sigma^*_\pm = \sigma_\pm^{-1}
$$

$$
[Q,\sigma^*_+\sigma_-] = [Q,\sigma^*_-\sigma_+] = 0
$$
$$
[Q_5,\sigma^*_+\sigma_-] = -2\sigma^*_+\sigma_-~;~ [Q_5,\sigma^*_-\sigma_+] = 2\sigma^*_-\sigma_+
$$
Without the zero-mode fields the model would be electrodynamics without charges.

The spurions (made from zero-mode operators) are the generators of large gauge transformations and are necessary to create a gauge invariant vacuum and therefore, to give a gauge invariant solution. We have
$$
     (\sigma_+^*\sigma_-)~ A_+~(\sigma_+\sigma_-^*) =A_+ + \fr{\sin{\kappa x^+}},{\kappa}
$$
So the vacuum must be an eigenstate of $\sigma_+^*\sigma_-$.  In fact, the physical vacuum is given by

$$
|\Omega (\theta )\rangle \equiv \sum_{M=-\infty }^{\infty }{\rm e}^{iM\theta
}|\Omega (M)\rangle \quad ;\quad |\Omega (M)\rangle =(-{\sigma }_{+}^{\ast }{
\sigma }_{-})^{M}|0\rangle
$$

This family of vectors are the only gauge invariant vectors which satisfy

\begin{equation}
\tilde{\Sigma}^{(+)}\ket{0} = \phi^{(+)}\ket{0} = \eta^{(+)}\ket{0} = Q_5\ket{0} = 0 \label{modevac}
\end{equation}

It is particularly easy to calculate the chiral condensate in light-cone gauge and obtain

$$
 \langle\Omega (\theta )|\bar{\Psi}\Psi|\Omega (\theta )\rangle = -{\frac{m},{2\pi}}
{\rm e}^\gamma \cos\theta
$$
The zero-mode fields also play an essential role in regulating the operator products in the theory.  To illustrate this fact consider the product of $\Psi_+$ with its conjugate

$$
    \langle \Psi_+^*(x+\epsilon)\Psi_+(x)\rangle\sim \fr{1},{2\pi\epsilon^-}
$$
This behavior allows us to define a gauge invariant Fermi product.  But it comes about through the combination

$$
   \langle :e^{i2\sqrt{\pi}\tilde{\Sigma}(x + \epsilon)}::e^{-i2\sqrt{\pi}\tilde{\Sigma}(x)}:\rangle \sim e^{-2\gamma}\fr{4},{m^2} \fr{1},{\epsilon^+\epsilon^-}
$$
with
$$
\langle e^{i2\sqrt{\pi}\eta^{(-)}(x+\epsilon)}\sigma_+^*e^{i2\sqrt{\pi}\eta^{(+)}(x+\epsilon)}e^{-i2\sqrt{\pi}\eta^{(-)}(x)}\sigma_+e^{-i2\sqrt{\pi}\eta^{(+)}(x)}\rangle\sim e^{\gamma}\kappa \epsilon^+
$$

Without the zero-mode field the operator product would be too singular to treat.  With it we can define

$$
\normord{\psi^\dagger\psi}=\lim_{\e\ra0\atop \e^2<0}\Biggl\{
	e^{-ie\int_x^{x+\e} A_\nu^{(-)}dx^\nu}
	\psi^\dagger(x+\e)\psi(x)e^{-ie\int_x^{x+\e} A_\nu^{(+)}dx^\nu}
		-{\rm V.E.V.}\Biggr\}
$$

Which allows the existence of the operator solution.  The zero-mode fields enter the solution in other, essential ways, but we shall now proceed 
to finding the solution by formulating the problem in the various quantization schemes.

\section*{Quantizing the Model}

Looking at the solution we see that it is straightforward to quantize the model on $t = 0$. The degrees of freedom are

$$
         \Psi_+~;~A_+~;~\Pi_A~;~\Psi_-
$$

The quantization follows completely standard lines.  We initialize $\psi_+$ and write it in the Bosonized basis.

$$
{\Psi}_+ = Z_+ {\rm e}^{{\Lambda}_+^{(-)}}{\sigma}_+ {\rm e}^
{{\Lambda}_+^{(+)}} 
$$

$$
\Lambda^{(+)}(0,x^1)=i(4\pi )^{-\ha}\int_{0}^{\infty
}dk_{1}k_{1}^{-1}c(k_{1})\left({\rm e}^{-ik_1 x^1}-\theta 
(\kappa -k_{1})\right)
$$

Looking at the solution we see that, in terms of the modes which diagonalize the Poincar\'e generators we have

$$
         c(k^1) = \eta^*(-k^1) + \sqrt{\fr{|k_1|},{\omega}}\left( \Sigma(k_1) + \Sigma^*(-k_1)\right)
$$
From (\ref{modevac}) we see that

$$
c(k^1)\ket{0_{ET}} = A_+^{(+)}\ket{0_{ET}} = \Pi_A^{(+)}\ket{0_{ET}}=0~~;~~\ket{0_{ET}} \neq \ket{0}
$$
Not only is the physical vacuum not the perturbative vacuum in the equal-time basis, it is extremely complicated, projecting on to every basis vector allowed by kinematics and the charge super selection rule.  In addition, the problem of finding the eigenstates of the Hamiltonian is a complicated dynamical problem.

We now consider the problem of quantizing the model on the surface $x^+ = 0$.  The degrees of freedom are

$$
       \Psi_+~;~\Psi_-~;~\lambda
$$

The main subtlety encountered in this quantization scheme is not the existence of the zero-mode fields.  If these fields are left out a solution cannot be found; but the fields are right there in the Lagrangian, the fact that they are zero-mode fields is easily learned from the equations of motion and the algebra of the zero mode fields can be determined by fairly straightforward procedures.  The most serious problem is that the Fermi product cannot be regulated by splitting in the $x^-$ direction.  There are various ways to get around the difficulty.  Here we will describe one which involves over regulating the theory, solving the dynamics then removing the over-regulation.

We first use the equations of motion to discover the dependence of $\Psi_+$ on $\eta$: 

$$
2{\partial }_{+}\partial_-A_+ = {\lambda } +m{\partial }_{+}{\phi }(x)-{\frac{m^{2}},{2}}A_{+}(x)
$$

$$
    \Longrightarrow A_+ \supset m\partial_+\eta
$$

$$
i{\partial}_{+}\Psi _{+}=e\Psi _{+}A_{+}
$$

$$
     \Longrightarrow \Psi _{+}={\rm e}^{-i2\sqrt{\pi }(\eta (x^{+}))}\Psi _{R}
$$
Where we have defined $\Psi _{R}$ simply to be the rest of $\Psi_+$. We find that $\Psi _{R}$ satisfies the canonical commutation relations on $x^+ = 0$ so we initialize it as 

$$
\{\Psi _{R}(x^{+},x^{-}),\Psi _{R}^{\ast }(x^{+},y^{-})\}={\delta }
(x^{-}-y^{-})
$$
$$
{\Psi }_{R}(0,x^{-})=Z{\rm exp}[-2i\sqrt{\pi }\tilde{{\phi }}^{(-)}(0,x^{-})]%
\tilde{{\sigma }}_{+}{\rm exp}[-2i\sqrt{\pi }\tilde{{\phi }}^{(+)}(0,x^{-})]
$$
$$
\tilde{{\sigma }}_{+}=exp\Big[i\sqrt{\pi }\{\tilde{Q}_{5}+\tilde{Q}%
\}(4m)^{-1}+\int_{0}^{\tilde{\kappa}}dk_{1}k_{0}^{-1}\{\tilde{c}(k_{1})-%
\tilde{c}^{\ast }(k_{1})\}\Big]
$$
$$
\tilde{\phi}^{(+)}(x^{-})=i(4\pi )^{-{\frac{1},{2}}}\int_{-\infty
}^{0}dk_{1}k_{0}^{-1}\tilde{c}(k_{1})\left( e^{-ik\cdot x}-\theta (\tilde{%
\kappa}-k_{0})\right)
$$
We now calculate $P_+$ 

$$
P_{+}= {\frac{m^2},{4}}{\int }_{-{\infty }}^{\infty }\tilde{\phi}^2 dx^{-} +{%
\int }_{-{\infty }}^{\infty }\{ ({\partial}_+{\phi})^2 -({\partial}_+{\eta}%
)^2 \}dx^{+}
$$

Since this operator is already diagonal, we now know the space-time dependence of $\Psi_+$.  As it stands the construction is not translationally invariant. But we can now define Fermi products by splitting in a space-like direction and recover covariance by taking the limit

$$
 \lim\tilde{\kappa} \rightarrow 0 \Longrightarrow \Psi_+ = Z_+ {\rm e}^{\Lambda_+^{(-)}}\sigma_+ {\rm e}^{\Lambda_+^{(+)}}
$$

We note that here, $\ket{0}$ is the perturbative vacuum so that the physical vacuum, $\ket{\Omega}$, is set by kinematics and gauge invariance.  In the same way we can discover that $\bra{\theta}\bar{\Psi}\Psi\ket{\theta} \neq 0$.  But we cannot calculate the value of the condensate from kinematics and gauge invariance: the value depends on the $\Psi_+$ wavefunction renormalization constant and that quantity is determined by maintaining the canonical commutation relation for $\Psi_+$ with the product defined by space-like splittings --- which requires a dynamical calculation.

We now turn to the problem of quantization on the surface $x^- = 0$ --- the anti-light-cone gauge.  The degrees of freedom are 
$$
         \Psi_-~;~A_+~;~\partial_-A_+
$$
A very unusual feature here is that not only is $\Psi_+$ not a degree of freedom, it is not a constraint which must be resolved prior to solving for the dynamics of the degrees of freedom.  Indeed, $\Psi_+$ must be solved for at the very end of the calculation, as we shall discuss below.  A very important feature of this formulation is that the $\Psi_-$ products can be defined by splitting in the initial value surface.  We can thus calculate

$$
J_{+}(x)=\lim_{y^{+}\rightarrow x^{+}}{\frac{e},{2}}\{{\psi }_{-}^{\ast
}(x^{+})\psi _{-}(y^{+}){\rm exp}[-ie%
\int_{y^{+}}^{x^{+}}A_{+}(z^+,x^-)dz^+]+h.c.\}
$$
$$
        J_{+}(x) =  m{\partial }_{+}{\phi }(x)-{\frac{m^{2}},{2}}A_{+}(x) 
$$
$$
   \lim_{y^{+}\rightarrow x^{+}}\{{\frac{i},{2}}{\psi}_{-}^{\ast}(x^+){
\partial }_{+}\psi _{-}(y^{+}){\rm exp}[-ie%
\int_{y^{+}}^{x^{+}}A_{+}(z,x^{-})dz]+h.c.-{\frac{1},{2{\pi }}}{\frac{1},{
(x^{+}-y^{+})^{2}}}\} 
$$
$$
   i{\Psi}_-^{\ast}{\partial}_+
{\Psi}_- = ({\partial }_{+}{\phi })^{2}-{\frac{m^{2}},{4}}(A_{+})^{2}
$$

Perhaps the main subtlety of the current formulation is not to be too quick to think that $A_+$ and $\partial_- A_+$ are canonical conjugates.  Indeed, although

$$
\lbrack A_{+}(x^{+},x^{-}),A_{+}(y^{+},x^{-})]=0
$$
and
$$
    [{\partial }_{-}A_{+}(x^{+},x^{-}),A_{+}(y^{+},x^{-})]=-{\frac{i},{2}}{
\delta } (x^{+}-y^{+})
$$
we have
$$
   [{\partial}_-A_+(x^+,x^-),{\partial}_-A_+(y^+,x^-)]=-i\frac{m^2},{16}
{\epsilon}(x^+-y^+)
$$
This problem is mearly a matter of being careful; the correct algebra can be obtained either by forcing agreement between the Heisenberg equations and the equations of motion or by the Dirac procedure.

Combining the last equation with

$$
   P_-=\int_{-{\infty }}^{\infty}({\partial}_-A_+)^2 dx^+
$$
gives

$$
{\partial }_-A_+(x)=-\frac{m},{2}\tilde{\Sigma}
$$

Again the dynamical operator is diagonal in the degrees of freedom.  From

$$
-2\partial_+\partial_-A_= = \lambda +   m{\partial }_{+}{\phi }(x)-{\frac{m^{2}},{2}}A_{+}
$$
we get

$$
A_+(x)=\frac{2},{m^2}\{ {\lambda}+m{\partial}_+({\phi}+\tilde{\Sigma}) \}
$$
or
$$
A_{+}={\frac{2},{m}}{\partial }_{+}(\tilde{{\Sigma }}+{\eta })
$$

We now turn to the problem of constructing $\Psi_+$. We do that through the equation

$$
i{\partial }_{+}\Psi _{+}=e\Psi _{+}A_{+}
$$
Formally, this equation is easy to solve.  In doing so we encounter some singular objects; but if we stick to the plan of regulating ultraviolet singularities by space-like point splitting, infrared singularities by a Klaiber subtraction and defining exponentials by Wick ordering, we recover the correct solution.

\end{document}